\documentclass[draftclsnofoot,onecolumn]{IEEEtran}
\IEEEoverridecommandlockouts
\usepackage{cite}
\usepackage{hyperref}
\usepackage{amsmath,amssymb,amsfonts}
\usepackage{amsmath}
\usepackage{amsthm}

\usepackage{graphicx}
\usepackage{textcomp}
\usepackage{xcolor}
\usepackage{graphicx}
\usepackage{float} 
\usepackage{subfigure}
\usepackage{amsmath}
\usepackage{amsfonts,amssymb}
\usepackage{mathrsfs}
\usepackage{mathtools}
\usepackage{algorithm}
\usepackage{algorithmicx}
\usepackage{algpseudocode}
\usepackage{bm}
\bibliography{IEEEabrv}
\def\BibTeX{{\rm B\kern-.05em{\sc i\kern-.025em b}\kern-.08em
    T\kern-.1667em\lower.7ex\hbox{E}\kern-.125emX}}

\begin{document}

\title{Mutual Information Approximation\\
\thanks{Supported by BUPT Excellent Ph.D. Students Foundation}
}

\author{Chongjun~Ouyang,~\IEEEmembership{Student Member,~IEEE}, 
        Sheng~Wu,
and Hongwen~Yang,~\IEEEmembership{Member,~IEEE}    
\thanks{C. Ouyang, S. Wu, and H. Yang are with the School of Information and Communication Engineering, Beijing University of Posts and Telecommunications, Beijing 100876, China. (E-mail: \{DragonAim, thuraya, yanghong\}@bupt.edu.cn)}
}

\maketitle

\begin{abstract}
To provide an efficient approach to characterize the input-output mutual information (MI) under additive white Gaussian noise (AWGN) channel, this short report fits the curves of exact MI under multilevel quadrature amplitude modulation ($M$-QAM) signal inputs via multi-exponential decay curve fitting (M-EDCF). Even though the definition expression for instanious MI versus Signal to Noise Ratio (SNR) is complex and the containing integral is intractable, our new developed fitting formula holds a neat and compact form, which possesses high precision as well as low complexity. Generally speaking, this approximation formula of MI can promote the research of performance analysis in practical communication system under discrete inputs.

\end{abstract}


\section{Introduction}
Mutual information (MI) represents the limit achievable transmission rate in communication system. In 1948, Shannon proposed the famous Shannon formula $\log_2\left(1+\gamma\right)$ to evaluate the transmission limitation under Gaussian inputs. Since then, researches on analysis of mutual information or channel capacity have become a hot topic in the field of telecommunication. Later, the concepts of secrecy capacity \cite{c1} and efficient capacity \cite{c2} were proposed in and , which further broaden the path of performance analysis and correlated optimization. Besides the mutual information analysis in additive white Gaussian noise (AWGN) channel, many theoretical results about the ergodic mutual information in wireless channel were also widely reported in the literature \cite{c3,c4,c5}. Nevertheless, it should be noticed that the inputs follow non-Gaussian distribution due to the limitation of digital modulation system. More specifically, the practical implementation of the communication system is based on finite alphabet inputs, and all the symbols are drawn form discrete constellation. Under this situation, it is impossible to adopt Shannon formula to employ the performance analysis. 

In 2007, Yang {\emph{et al.}} in \cite{c6} formulated a series expression for the ergodic mutual information under BPSK modulation mode over Nakagami-$m$ fading. Although the contribution in \cite{c6} is seemingly fascinating, its constraint by limiting $m$ as an integer indeed makes it lack generality. Besides, the proposed formula is hard computable. More serious, existing literature failed to offer an analytical framework to analyze the mutual information under discrete inputs. In 2007, the authors in \cite{c7} and \cite{c8} developed some approximation formulas for the instantaneous mutual information under AWGN channels. Although these formulas hold compact forms, the approximation error was relatively high. Motivated by these work, we will also develop some approximation expression for the mutual information under discrete inputs. 

The remaining parts of this manuscript is structured as follows: Section \ref{section2} presents the approximated expression for the mutual information. In Section \ref{section3}, simulation results are provided. Section \ref{section4} offer some further discussions of our work. Finally, Section \ref{section5} concludes the paper.

\section{Approximation of Mutual Information}
\label{section2}
Consider a single-input single-output AWGN channel under finite alphabet inputs, the input-output mutual information can be expressed \cite{b1}
\begin{equation}
\label{eq1}
I\left(X;Y\right)=\sum_{k=0}^{m-1}p(k)\int_{-\infty}^{+\infty}p\left(y|x_k\right)\log_2\left(\frac{p\left(y|x_k\right)}{\sum\limits_{i=0}^{m-1}p(i)p\left(y|x_k\right)}\right){\rm{d}}y.
\end{equation}
Here, we assume the input signal utilizes the $M$-QAM modulation and $x_k\in{\mathcal{X}}=\{x_1,x_2,\cdots,x_M\}$. Equ. \eqref{eq1} can be also rewritten as
\begin{equation}
\label{eq2}
I\left(X;Y\right)=\sum_{k=0}^{m-1}p(k)\mathbb{E}\left[\log_2\left(\frac{p\left(y|x_k\right)}{\sum\limits_{i=0}^{m-1}p(i)p\left(y|x_k\right)}\right)\right]=\mathbb{E}\left[\sum_{k=0}^{m-1}p(k)\log_2\left(\frac{p\left(y|x_k\right)}{\sum\limits_{i=0}^{m-1}p(i)p\left(y|x_k\right)}\right)\right].
\end{equation}
On the basis of Equ. \eqref{eq2}, the mutual information can be obtained by Monte-Carlo simulation, which is really time-consuming. On the other hand, the integral in Equ. \eqref{eq1} is intractable. Due to these challenges, we will not continue to try to simplify the integrals with any obscure mathematical theories but to approximate the mutual information directly by curve fitting. From this perspective, we may figure out some elegant approximate expressions of the mutual information, with both accurate precision and compact form. 

In the following part, some classical $M$-QAM modulation modes will be considered, including 4-QAM, 16-QAM, 64-QAM and 256-QAM. There are many effective tools for curve fitting, such as Matlab, OriginPro, SAS, and SPSS. In this manuscript, a classical software named 1stQpt is utilized, which fits the curves on the basis of Levenberg-Marquardt(LM)  and  Universal  Global  Optimization(UGO)  algorithms. Actually, for different modulation mode, they can all be approximated as the following form
\begin{equation}
{\mathcal{I}}_{M-{\rm{QAM}}}\left(\gamma\right)=\left(\log_2M\right)\left(1-\sum_{k=1}^{N}a_i\exp\left(-b_i\gamma\right)\right),
\end{equation} 
and $\gamma$ denotes the SNR. As can be seen form this equation, our approximation is based on multi-exponential decay curve fitting (M-EDCF). After the experiments, the coefficient for different modulation mode are summarized in Table 1. Notice that the RMSE in this table means Root Mean Square Error (RMSE), i.e. the standard deviation of the residuals (prediction errors).
\begin{table}[htbp]
\centering
\label{table1}
\caption{Fitting Coefficients}
\begin{tabular}{ccccccccc}
\hline\hline
\multicolumn{9}{c}{256-QAM}                                                                                                   \\ \hline\hline
$a_1$          & $a_2$          & $a_3$          & $a_4$          & $b_1$          & $b_2$          & $b_3$          & $b_4$          & RMSE        \\ \hline
0.228768   & 0.229083 & 0.118223 & 0.423927   & 0.183242 & 0.038011 & 0.994472 & 0.006911 & 0.000592021 \\
0.329121 & 0.197647 & 0.473233 &             & 0.074934 & 0.662915 & 0.007544 &             & 0.002444037 \\ \hline\hline
\multicolumn{9}{c}{64-QAM}                                                                                                    \\ \hline\hline
$a_1$          & $a_2$          & $a_3$          & $a_4$          & $b_1$          & $b_2$          & $b_3$          & $b_4$          & RMSE        \\ \hline
0.198324 & 0.512831 & 0.209086 & 0.079759 & 0.408618 & 0.027517 & 0.120616 & 1.467118  & 0.000238369 \\
0.545415  & 0.15446 & 0.300125 &             & 0.028609 & 1.016175 & 0.191244 &             & 0.000815762 \\
0.360522 & 0.639478 &             &             & 0.475473 & 0.033426 &             &             & 0.006181322 \\ \hline\hline
\multicolumn{9}{c}{16-QAM}                                                                                                    \\ \hline\hline
$a_1$          & $a_2$          & $a_3$          & $a_4$          & $b_1$          & $b_2$          & $b_3$          & $b_4$          & RMSE        \\ \hline
0.658747 & 0.117219 & 0.224034 &             & 0.115521 & 1.467927 & 0.482023  &             & 0.000318074 \\
0.277888 & 0.722112 &             &             & 0.898478 & 0.123144 &             &             & 0.001629419 \\ \hline\hline
\multicolumn{9}{c}{4-QAM}                                                                                                    \\ \hline\hline
$a_1$          & $a_2$          & $a_3$          & $a_4$          & $b_1$          & $b_2$          & $b_3$          & $b_4$          & RMSE        \\ \hline
0.143281 & 0.856719 &             &             & 1.557531 & 0.57239  &             &             & 0.00036472  \\
1           &             &             &             & 0.6507      &             &             &             & 0.0034287 \\ \hline 
\end{tabular}
\end{table}

\section{Numerical Results}
\label{section3}
In this section, we use simulation results to verify the precision of the proposed approximate expressions. As can be seen from these graphs, our proposed approximation holds high precision, which can be used to evaluate the mutual information. It should be noted that we only use the approximation with the smallest RMSE to do the estimation. 
\begin{figure}[htbp] 
    \centering
    \subfigure[4-QAM]
    {
        \includegraphics[width=0.40\textwidth]{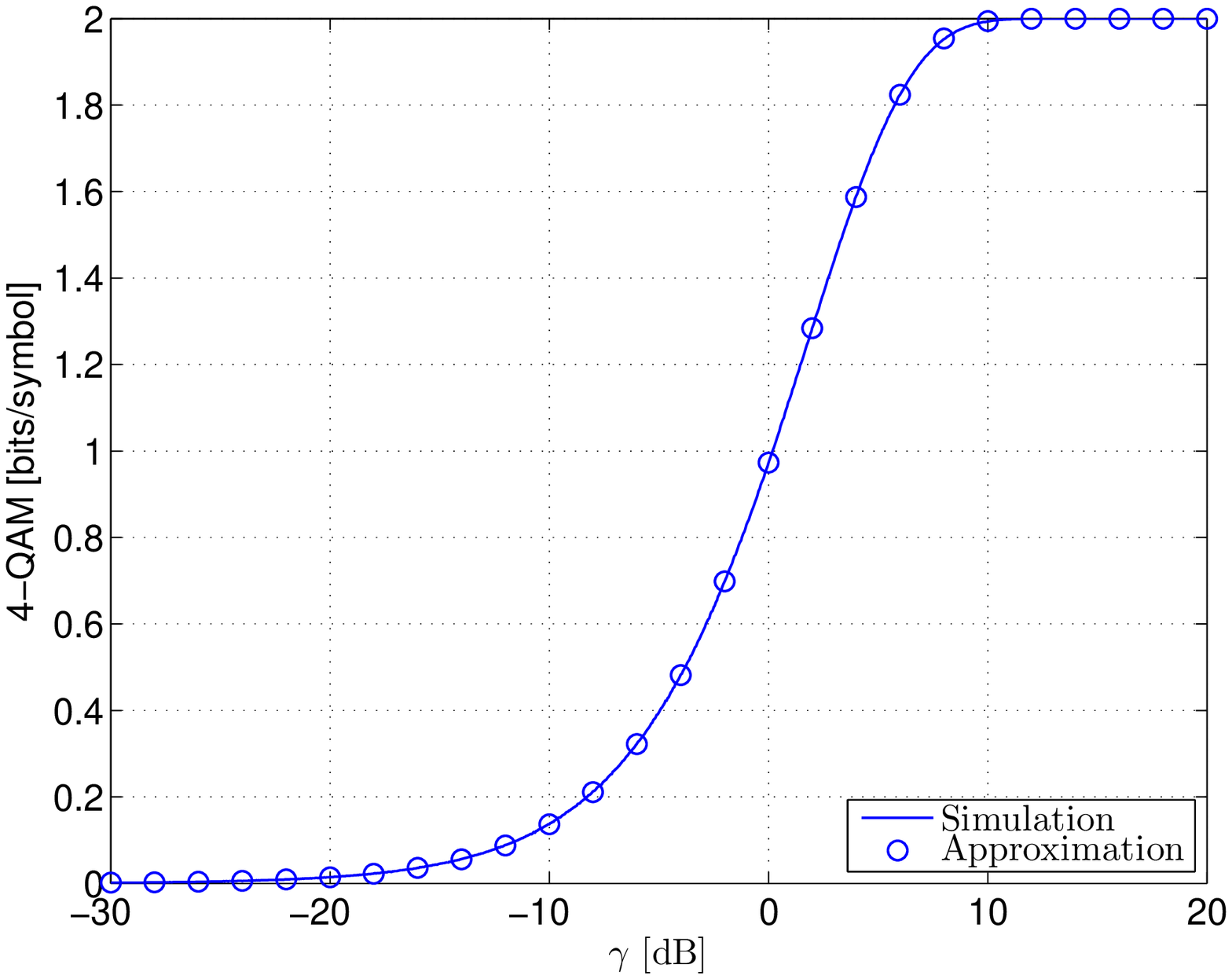}
	   \label{fig1a}	   
    } 
\vspace{-5pt}
\hspace{-15pt}
   \subfigure[16-QAM]
    {
        \includegraphics[width=0.40\textwidth]{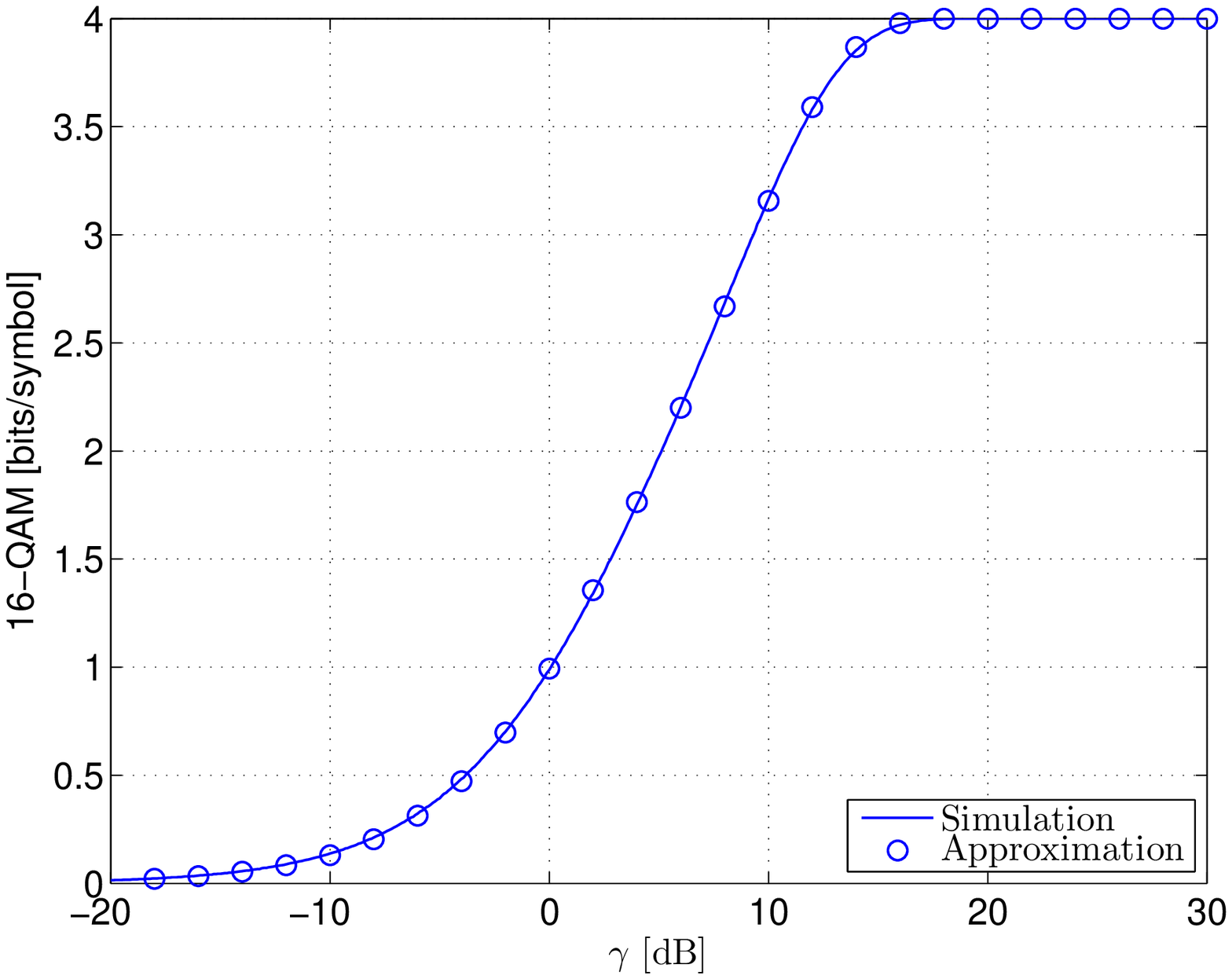}
	   \label{fig1b}	   
    } \\
    \subfigure[64-QAM]
    {
        \includegraphics[width=0.40\textwidth]{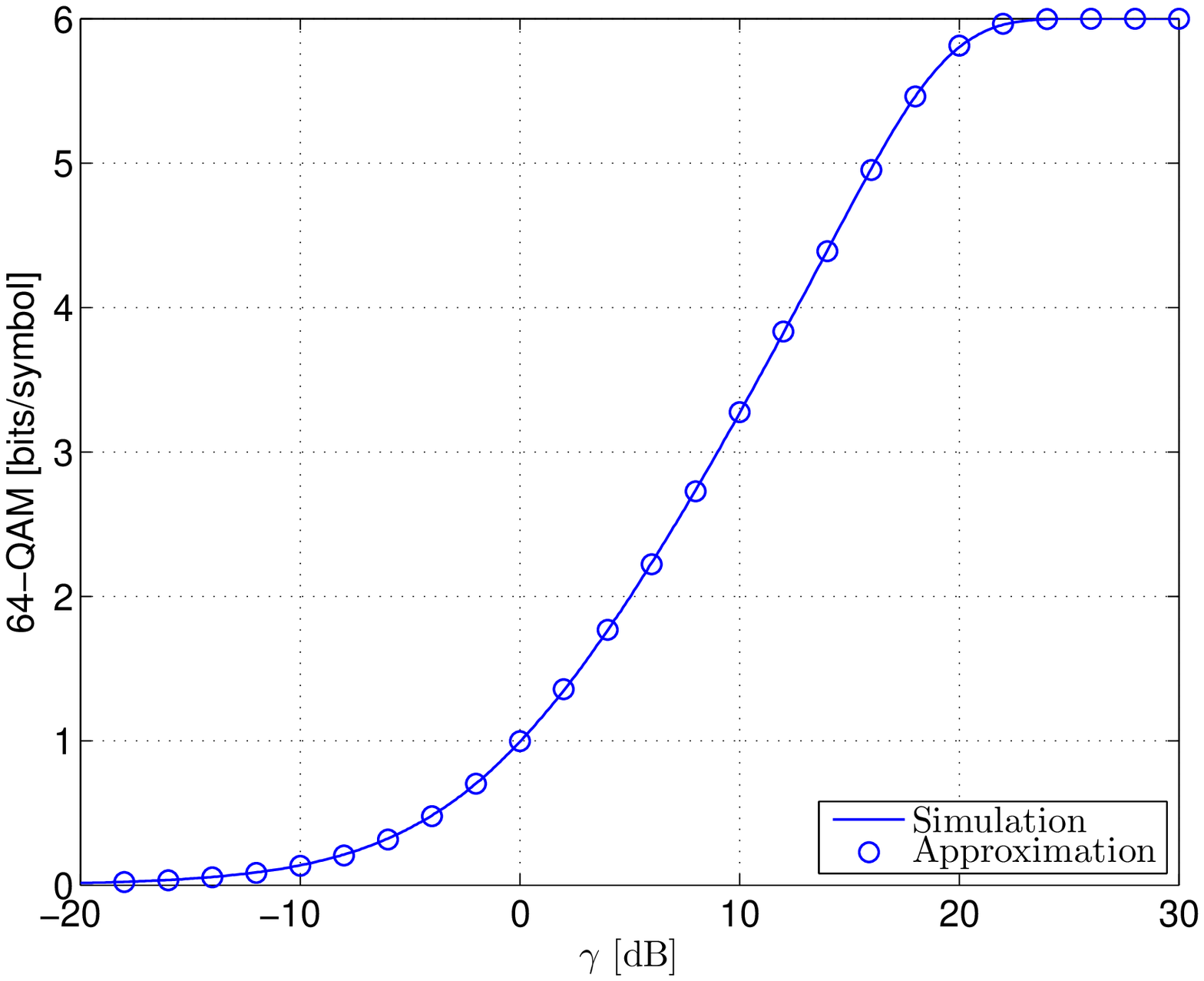}
	   \label{fig1c}	   
    } 
\vspace{-5pt}
\hspace{-15pt}    
    \subfigure[256-QAM]
    {
        \includegraphics[width=0.40\textwidth]{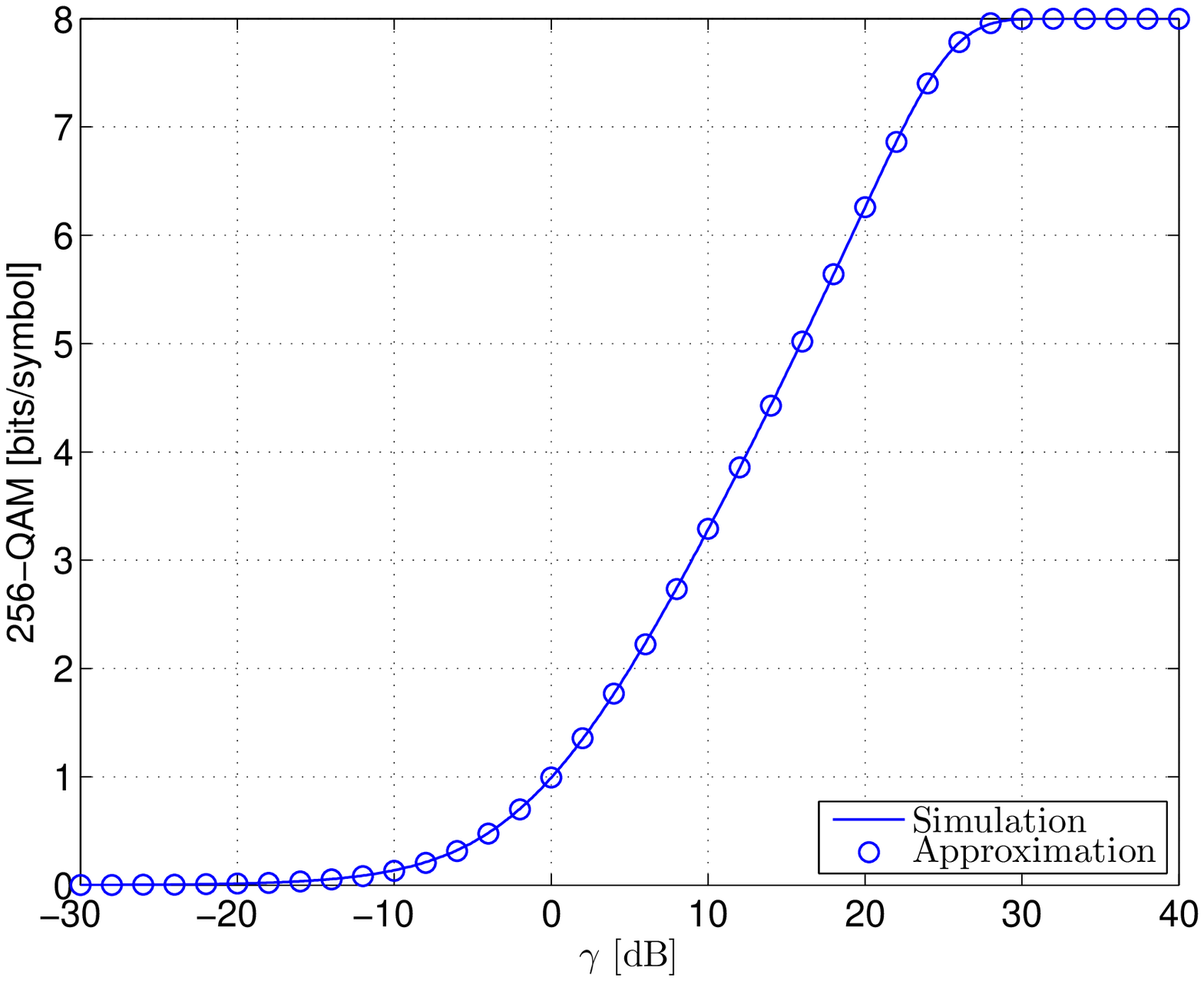}
      \label{fig1d}	        
    }
\\
    \caption{Simulated and approximated ergodic mutual information for different modulation modes. }
    \label{figure1}
	\vspace{0.2in}
\end{figure}

\section{Further Discussion}
\label{section4}
Let us move back to the proposed approximation formula which reads:
\begin{equation}
{\mathcal{I}}_{M-{\rm{QAM}}}\left(\gamma\right)=\left(\log_2M\right)\left(1-\sum_{k=1}^{N}a_i\exp\left(-b_i\gamma\right)\right).
\end{equation}
Indeed, the form of this formula is considerably neat and simple. Besides, as can be seen from the simulation results, it provides a robust approximation to the exact mutual information. Due to its high precision and neat form, this expression can be utilized to analyze the performance of all kinds of wireless channel. For example, the ergodic information under any wireless channel with PDF $f\left(\gamma\right)$ can be written as:
\begin{equation}
\int_0^{\infty}{\mathcal{I}}_{M-{\rm{QAM}}}\left(\gamma\right)f\left(\gamma\right){\rm{d}}\gamma=\left(\log_2M\right)\left(1-\sum_{k=1}^{N}a_i\int_0^{\infty}\exp\left(-b_i\gamma\right)f\left(\gamma\right){\rm{d}}\gamma\right).
\end{equation}
As it shows, the core point of this calculation is to calculate the moment generating function of $f\left(\gamma\right)$. In the future work, we will try to use this formula to analyze the ergodic mutual information, the secrecy capacity and the efficient capacity under finite alphabet inputs.

\section{Conclusion}
\label{section5}
This paper provides a unified and general framework to analyze the mutual information under finite alphabet inputs. Actually, it has transformed an intractable problem into a relatively solvable one, i.e., the mutual information analysis under discrete inputs. Even though these derivations are derived on the basis of numerical fitting, it really offers a fast, efficient and precise method to solve these problem.

\vspace{12pt}
\end{document}